\documentstyle[aps,pra,twocolumn,epsfig,amsmath,amssymb,graphics]{revtex}

\begin{document}
\title{Multipartite Classical and Quantum Secrecy Monotones}
\author{N. J. Cerf,$^1$ S. Massar,$^2$ and S. Schneider$^{3}$}
\address{$^1$ Ecole Polytechnique, CP 165, Universit\'e Libre de Bruxelles,
1050 Brussels, Belgium\\
$^2$ Service de Physique Th\'eorique, CP 225, Universit\'e Libre
de Bruxelles, 1050 Brussels, Belgium\\
$^3$ Department of Chemistry, University of Toronto,
Toronto, ON, M5S 3H6, Canada}

\date{February 2002}

\maketitle

\begin{abstract}
In order to study multipartite quantum cryptography, we introduce
quantities which vanish on product probability distributions, and
which can only decrease if the parties carry out local
operations or carry out public classical communication. These
``secrecy monotones'' therefore measure how much secret correlations
are shared by the parties. In the bipartite case we show that the
mutual information is a secrecy monotone. In the multipartite case we
describe two different generalisations of the mutual information, both
of which are secrecy monotones. The existence of two distinct secrecy
monotones allows us to show that in
multipartite quantum cryptography the parties must make irreversible
choices about which multipartite correlations  they want to
obtain. Secrecy monotones can be extended to the quantum domain
and are then
defined on density matrices. We illustrate this generalisation
by considering tri-partite quantum cryptography based on the
Greenberger-Horne-Zeilinger (GHZ)
state. We show that before carrying out measurements on the state, the
parties must make an irreversible decision about what probability
distribution they want to obtain.
\end{abstract}

%\pacs{PACS numbers: 03.67.-a, 03.65.Bz, 42.50.Dv}

\section{Introduction}

Quantum cryptography uses the
uncertainty principle of quantum mechanics to allow two parties
to communicate secretly\cite{BB}. It has been
extensively studied during the last decade 
both theoretically and experimentally (see e.g. \cite{GRTZ} for a review). 
The basic idea of a quantum cryptographic protocol
is that two parties, Alice and Bob, use a quantum communication channel to
exchange an entangled state $\Psi_{AB}$. Local measurements yield correlated
results and allow them to obtain a certain number of shared secret bits
$P^2_{AB}$ uncorrelated with Eve, defined by the probability distribution
$P^2_{AB}(0,0)=P^2_{AB}(1,1)=1/2$. These
resulting secret bits can then be used for secure cryptography,
using e.g. the one-time pad scheme. An eavesdropper, Eve, will necessarily
disturb the quantum state $\Psi_{AB}$ when attempting to get information
on the secret bits, and will therefore be detected by
Alice and Bob. A central problem of quantum cryptography is to
establish the maximum rate at which Alice and Bob can establish a
secret key for a given level of disturbance by Eve.

Independently of quantum cryptography,
Maurer has introduced a paradigm for classical cryptography based on
probabilistic correlations between two parties, Alice and Bob, and an
eavesdropper Eve\cite{Maurer}.
 Specifically, suppose that Alice, Bob, and Eve
have several independent realizations of their three random variables,
distributed each according to the probability distribution $P_{ABE}$.
The goal is for Alice and Bob to distill, from these realizations,
a maximal number of shared secret bits $P^2_{AB}$.
The tools available to Alice and Bob to perform this distillation
are local operations and classical public communications (LOCC).
An important question is to estimate the maximum rate
at which the parties can generate the secret bits. Bounds on this
secret bits distillation rate have been
obtained in \cite{Maurer:Wolf2}.

Quantum cryptography and Maurer's cryptographic paradigm are closely
related. Indeed,
after measuring their quantum bits, the parties A, B and E end up 
in exactly the situation considered by Maurer. Therefore,
distillation protocols used in Maurer's cryptographic scheme can be
adapted to the quantum situation\cite{Gisin:Wolf1}. Moreover,
ideas from quantum cryptography and quantum information
theory have illuminated the structure of Maurer's cryptographic
scheme\cite{Gisin:Wolf2}.

In this paper, we will consider multipartite cryptography both from the
point of view of Maurer's classical cryptographic scheme and from the point of
view of quantum cryptography. As in the papers mentioned above, these
two approaches are complementary. We show that putting ideas from these two
approaches together provides new insights into multipartite cryptography.
Let us first consider the extension of Maurer's cryptographic
scenario to more than two parties.
One supposes that the different parties $A_1, A_2 \ldots A_n, E$
possess independent realizations
of $n+1$ random variables distributed according to the multipartite
probability distribution $P_{A_1 A_2\ldots A_n E}$ where $E$
denotes the eavesdropper, as before. In this case, however,
it is not obvious what the goal of the parties should be. For instance, in
the case of three parties, one possible aim of the distillation process
could be to maximize the resulting
number of random bits shared between pairs of parties
$P^2_{AB} P^2_{BC}P^2_{CA}$. Another goal could be to generate efficiently 
the tripartite probability distribution $P^3$ defined by $P^3_{ABC}(0,0,0)=
P^3_{ABC}(1,1,1) = 1/2$. This probability distribution allows any
one of the parties to encrypt a message (by {\sc xor}ing it with his random
variable and publicly communicating the result)
in such a way that it can be decrypted by both the other parties
independently. A third possibility could be
to generate the probability distribution
$P^x_{ABC}$ in which any two random variables are
independent random bits, while the third one is the {\sc xor} of
the other two bits,
$P^x_{ABC}(0,0,0) = P^x_{ABC}(1,1,0) =
P^x_{ABC}(1,0,1) = P^x_{ABC}(0,1,1) = 1/4$. This
probability distribution allows any one of the parties to encrypt a
message (by {\sc xor}ing it with his random variable and publicly
communicating the result)  in such a way that it can only be
decrypted if the other two parties get together and compute the {\sc xor}
of their two random bits.

One main result of the present paper is to show that in the
multipartite case, the parties must decide 
before they start the distillation protocol which probability
distribution they want to obtain. Making the wrong choice entails an
irreversible loss. For instance, the parties will in general obtain
more triplets $P^3$ if they directly distill to $P^3$
than if they first distill to another kind of probability
distribution, say shared
random bits between pairs of parties $P^2_{AB} P^2_{BC}P^2_{CA}$, and
 subsequently try to generate $P^3$ from these pairs of random bits.
More generally, we address in this paper the question of the
convertibility, using local operations and public communication, of
one multipartite probability distribution into another one.

As mentioned above, these issues can be generalized to
quantum-mechanical systems in the context of quantum cryptography. 
We thus aim at addressing the same questions of the convertibility between
quantum multipartite density operators in this paper. An important potential
application of this extension is to provide bounds on the yields
of multi-partite quantum cryptography. Indeed, in quantum
cryptography, the parties start with a quantum state (in general a
mixed state) and, by carrying out local operations, measurements,
and classical communication, they aim at obtaining a multi-partite
classical probability distribution (which can of course be viewed
as a particular kind of mixed quantum state). Bounds on the
interconvertibility of multipartite quantum states have been
studied by several authors (see for
instance\cite{BPRST,LPSW,GPV,PV}), but most of this work has focused on
pure states. The interconvertibility of mixed states,
and, in particular, applications to multipartite
cryptography, have so far been relatively little studied.
As an illustration, we consider in detail the case of quantum
cryptography based on the GHZ state $\Psi_{GHZ} = (|000\rangle +
|111\rangle)/\sqrt{2}$. By measuring $\Psi_{GHZ}$ in the $z$ basis the
parties can obtain the probability distribution $P^3$, while by
measuring in the $x$ basis, they can obtain the probability
distribution $P^x$. We show that the parties cannot obtain more
than one $P^3$ or one $P^x$ distribution per GHZ state. Combining
this with the bounds stated above, we see that when extracting
correlations from the GHZ state, the parties must make
an irreversible choice of what kind of tripartite correlations
they want to obtain.

In order to study these interconvertibility issues,
we have developed a tool which we call {\em secrecy monotones}. These
are functions of the multipartite probability distributions 
(or, more generally, of the quantum density operators) 
that can only {\em  decrease}
under local operations and public classical communication.
Therefore, comparing the value of the monotone on the initial and the
target probability distribution allows one to obtain an upper bound on the
number of realizations of the target probability distribution 
that can be obtained from the initial probability distribution. 
In fact, the upper bounds obtained in
\cite{Maurer} and in \cite{Maurer:Wolf2} on the secret key
distillation rate in the bipartite case
can be reexpressed in terms of the existence of certain
bipartite secrecy monotones.

Monotones have proved to be extremely useful
for the study of quantum entanglement (see for instance \cite{HHH}),
in which context they are called entanglement monotones. 
Our study of secrecy monotones is closely
inspired by these works on entanglement monotones. 
Entanglement monotones are positive and vanish on unentangled
density matrices, which implies that they measure the amount of entanglement in
a density matrix. In a similar way, secrecy monotones are positive
and vanish on product probability distributions
(or product density operators), so that they measure the
amount of both classical and quantum correlations between the parties.
In the present paper, we introduce two information-theoretic multipartite
secrecy monotones (called $S_n$ and $T_n$), which can be viewed as the
multipartite extension of the (classical or quantum) mutual information
of a bipartite system. The definition of the quantum mutual
information was discussed in \cite{ca1}, and its use in the context
of quantum channels was investigated in details in \cite{ca2}.
Here, this quantity is shown to be a monotone, and extended to 
multipartite systems. In particular, we discuss several applications 
of these monotones to the special case of three parties.

The paper is organised as follows. The first sections of the paper are
devoted exclusively to the classical secrecy monotones.
We begin in section \ref{general}
by giving a general definition of classical secrecy monotones and
studying the implications of this definition.
In section \ref{twomono} we introduce two specific multipartite
secrecy monotones $S_n$ and $T_n$ which are the multipartite
generalization of the bipartite mutual entropy.
Most of this section is devoted to proving that these functions obey
all the properties of a secrecy monotone. In section \ref{tri},
we use these two secrecy monotones to study the particular case of
tri-partite cryptography. In particular we address the question
raised above concerning the interconvertibility of the
probability distributions $P^2_{AB}$, $P^2_{BC}$, $P^2_{CA}$, $P^3_{ABC}$,
and $P^x_{ABC}$.  Finally, in section \ref{quantum}, we study the
generalization of the classical secrecy monotones to quantum
mechanis. In particular, we show that the monotones $S_n$ and $T_n$ have 
natural quantum analogues that have important applications
to multipartite quantum cryptography. As an illustration, we
study bounds on quantum cryptography based on the GHZ state.

\section{Properties of secrecy monotones}\label{general}

\subsection{Defining properties}\label{deff}

A secrecy monotone is a function $M$ defined on multipartite probability
distributions $P_{A_1 A_2\ldots A_n E}$ which obeys a series of
properties which we now review and explain. (We restrict ourselves to
the classical case, the quantum case will be analyzed 
in section \ref{quantum}.)
We will denote the monotone either $M(P_{A_1 A_2\ldots A_n E})$ or
$M(A1{\rm:}A2{\rm:}\ldots{\rm:}A_n{\rm:}E)$ where the semicolons separate the
different
parties.
\\

The first two properties ensure that $M$ provides a measure of
the amount of correlations between the parties.

1) {\em Semi-positivity:}
\begin{equation}
M(P_{A_1 A_2\ldots A_n E})\geq 0
\end{equation}

2) {\em Vanishing on product probability distributions:}
\begin{eqnarray}
&\mbox{if }&
 P_{A_1 A_2\ldots A_n E} = P_{A_1 E} P_{A_2 E} \ldots P_{A_n E}\ ,
\nonumber\\
&\mbox{then }&
M(P_{A_1 E} P_{A_2 E} \ldots P_{A_n E}) =0\ .
\end{eqnarray}

The next two properties express the monotonicity of
$M$ under LOCC, namely the fact $M$ can only decrease if one of the parties
performs some local operation (e.g. randomization) or publicly discloses
(partly or completely) the value of his variable. Thus, these 
monotonicity properties imply that $M$ describes the amount of correlations not
shared with Eve.  They also make $M$  useful for studying the
convertibility of one probability distribution into another.

3) {\em Monotonicity under local operations.} \\
Suppose that party $A_j$
carries out a local transformation that modifies $A_j$ to $\bar{A}_j$
according to the
conditional probability distribution $P_{\bar{A}_j | A_j}$. Then $M$
can only decrease:
\begin{eqnarray}
&\mbox{if }&
P_{A_1 \ldots \bar{A}_j\ldots A_n E} =
P_{\bar{A}_j | A_j} P_{A_1 \ldots {A}_j\ldots A_n E}\ ,
\nonumber\\
&\mbox{then }&
M(P_{A_1 \ldots \bar{A}_j\ldots A_n E}) \leq
M(P_{A_1 \ldots {A}_j\ldots A_n E})\ .
\end{eqnarray}

4) {\em Monotonicity under public communication.} \\
Suppose that party
$j$ publicly discloses the value of $\bar{A}_j$, where $\bar{A}_j$
depends on $j$'s variable $A_j$ according to the conditional probability
distribution $P_{\bar{A}_j|A_j}$. Then $M$ can only decrease:
\begin{equation}
M(P_{A_1 \ldots {A}_j\ldots A_n E| \bar{A}_j})
\leq M(P_{A_1 \ldots {A}_j\ldots A_n E})\ .\end{equation}

The next two properties are important if the secrecy monotone is to
provide information on the asymptotic rate of convertibility of one
probability distribution into another. By this, we mean that the
parties initially have a large number $n$ of realizations of the
probability distribution $P^1$ and want to obtain a large number $m$
of realizations of the probability distribution $P^2$. Property 5 ensures
that one can use the monotone $M$ to study the asymptotic limit $n,m \to
\infty$. Property 6 allows one to study the situation where one does
not want to obtain the exact probability distribution $(P^2)^{\otimes m}$,
but only a probability distribution that is close to
$(P^2)^{\otimes m}$.

5) {\em Additivity:} 
\begin{equation}
M(P^1 \otimes P^2) = M(P^1) + M(P^2)
\end{equation} 
Note that one may also impose only the weaker condition
$M(P^{\otimes n}) = n M(P)$ (see \cite{HHH} for a motivation for
considering only this weaker condition in the case of entanglement).

6) {\em Continuity:} \\
$M(P)$ is a continuous function of the probability
distribution $P$. We will not make more explicit the condition that
this imposes on $M$ since the monotones we will explicitly describe
below are highly smooth functions of $P$. 
We refer to \cite{HHH} where a weak continuity
condition is introduced and motivated in the context of entanglement.
\\

Finally, we introduce two additional properties which are
natural to impose if the monotone is to measure the amount of secrecy
shared by the parties $A_1 \ldots A_n$, with $E$ viewed as a hostile
party. Indeed, these final properties express the fact that the secrecy 
can only increase if $E$ looses information either by performing some
local operation or by publicly disclosing (in part or totally) his variable.

7) {\em Monotonicity under local operations by Eve.}\\
Suppose that Eve
carries out a local transformation which modifies $E$ to $\bar{E}$
according to the
conditional probability distribution $P_{\bar{E} | E}$. Then $M$
can only increase:
\begin{eqnarray}
&\mbox{if }&P_{A_1 \ldots A_n \bar {E}} =
P_{\bar{E} | E} P_{A_1 \ldots A_n E}\ ,
\nonumber\\
&\mbox{then }&
M(P_{A_1 \ldots A_n \bar {E}}) \geq
M(P_{A_1 \ldots A_n E})\ .\end{eqnarray}

8) {\em Monotonicity under public communication by Eve.} \\
Suppose that Eve publicly discloses the value of $\bar{E}$, where $\bar{E}$
depends on Eve's variable $E$ according to the conditional probability
distribution $P_{\bar{E}|E}$. Then $M$ can only increase:
\begin{equation}
M(P_{A_1 \ldots A_n E| \bar{E}})
\geq M(P_{A_1 \ldots A_n E})\ .
\end{equation}

\subsection{Consequences of the defining properties}

\subsubsection{Upper bound on the yield}

The most important consequence of the defining properties is that a
monotone allows one to obtain a bound on the rate at which a
multipartite probability distribution $P^1$ can be converted into another
probability distribution $P^2$ using LOCC. Suppose that the parties
are able, using LOCC, to convert $n$ realizations of $P^1$ into some realization 
of a probability distribution $P^{2'}$ which is close to $m$ independent
realizations of the desired
probability distribution $P^2$:
\begin{equation}
(P^1)^{\otimes n} {\longrightarrow \atop LOCC} P^{2'} \simeq
(P^2)^{\otimes m} \ .
\label{convert}
\end{equation}
The yield of this distillation protocol is defined as
\begin{equation}
Y_{P^1 \to P^2} = {m \over n} \ .
\end{equation}
The existence of a secrecy monotone M allows us to put a bound on the
yield. Indeed, from Eq.~(\ref{convert}), we have
\begin{equation}
M((P^1)^{\otimes n}) = n M(P^1) \geq M(P^{2'}) \simeq m M(P^2)
\end{equation}
where we have used the defining properties of M (additivity,
monotonicity and continuity).
Hence, using the positivity of $M$, we obtain
\begin{equation}
Y_{P^1 \to P^2} \leq { M(P^1) \over M(P^2) } \ .\end{equation}

\subsubsection{Monotones that do not involve Eve}\label{NoEve}

In practice, it is often much easier to construct a restricted type of
monotones $M$ that are
only defined on probability distributions $P_{A_1 A_2\ldots A_n}$ that do
not depend on $E$. These simple monotones are therefore applicable
only to the cases where Eve initially has no
information about the probability distribution.
Importantly, one can
easily extend such monotones $M$ to more general monotones $M$ defined on
probability distributions  $P_{A_1 A_2\ldots A_n  E}$ 
that also include initial correlations with Eve.
The simplest way to carry out this
extension is to calculate the probability distributions 
$P_{A_1 A_2\ldots A_n |  E}$ conditional on Eve's variable $E$,
and then to average the values of $M$ 
on the conditional probability distribution. 
This yields a monotone $M_1$:
$$
M_1(P_{A_1 A_2\ldots A_n  E}) = \sum_E P(E) M(P_{A_1 A_2\ldots A_n |
  E})\ .
$$
The monotone $M_1$ thus constructed obeys property 8, but in general does
not obey property 7\cite{Maurer:Wolf2}.

In order to obtain a monotone that obeys both properties 7 and 8, before computing
the conditional probability distribution $P_{A_1 A_2\ldots A_n |  E}$, 
we first need to take the minimum over Eve's operations. This
transforms the variable $E$ into $\bar E$ according to $P(\bar{E} | E)$. 
This procedure yields a new monotone $M_\downarrow$:
$$
M_\downarrow (P_{A_1 A_2\ldots A_n  E}) = \min_{P(\bar{E} | E)}
\sum_{\bar{E}} P(\bar{E}) M_|(P_{A_1 A_2\ldots A_n |  \bar{E}})
\ .
$$
Note that it is this second procedure that was used in
\cite{Maurer:Wolf2} to obtain a strong upper bound on the rate of
distillation of a secret key.

\subsubsection{Extending monotones to more parties}

A monotone defined on a $n$-partite
probability distribution can be extended in a natural way to a monotone on a
$m$-partite probability distribution with $m>n$.
Let us illustrate this procedure in
the case of a bipartite monotone $M_2(A{\rm:}B)$ extended to a
tripartite case. A tripartite monotone for the variables
$A$, $B$, and $C$ is simply
$M_2(AB{\rm:}C)$ and can be interpreted as the bipartite monotone which
would be obtained if parties $A$ and $B$ get together. We can of course
group the parties in many different ways, and therefore $M_2(AC{\rm:}B)$
and $M_2(BC{\rm:}A)$ are two other independent tripartite monotones.
These three monotones are distinct from the genuinely tripartite
monotones that can be constructed on $P_{ABC}$, as we will show later on, 
and lead to independent conditions on the convertibility
of tripartite distributions.

\section{Two classical multipartite secrecy monotones}\label{twomono}

We now introduce two information-theoretic
multipartite secrecy monotones for $n$ parties 
$A_1,\ldots A_n$ (with $n\ge 2$)
sharing some classical probability distribution $P_{A_1 \cdots A_n}$.
We shall suppose that Eve initally has no knowledge about the
probability distribution. The generalization to the case where the
probability distribution depends on $E$ can be done as shown in
section \ref{NoEve}.

\subsection{Amount of shared randomness between the parties: $S_n$}

The first multipartite secrecy monotone is denoted $S_n$ and
defined by
\begin{eqnarray}
& & S_n (A_1 {\rm:} \cdots {\rm:} A_n)  =
H(A_1\cdots A_n) \nonumber \\
& &\quad \quad \quad -\sum_{i=1}^n H(A_i|A_1\cdots A_{i-1}A_{i+1}\cdots
A_n)
\label{SSS1}
\end{eqnarray}
where  $H(A)$ denotes the Shannon entropy of variable $A$ distributed as
$P_A$, that is, $H(A)= - \sum_a p_a \log p_a$.

In order to provide a physical interpretation to $S_n$, we note that
the first term on the right hand side of Eq. (\ref{SSS1})
is the total randomness of the probability distribution 
$P_{A_1 \cdots A_n}$ whereas the subtracted terms 
are the amounts of randomness that
are purely local to each party. Thus $S_n$
measures the number of bits of shared randomness between
the $n$ parties (irrespective of them being shared between two, three, or more
parties, but not including the local randomness).
On the basis of this interpretation it is natural that if one of the
parties publicly reveals some of his data, this will decrease $S_n$
since the total number of bits of shared randomness has
decreased. This remark suggests
that $S_n$ should be a secrecy monotone. That this is indeed the case
will be proven below.

We begin by introducing two alternative expressions for $S_n$:
\begin{eqnarray}
\lefteqn{S_n (A_1 {\rm:} \cdots {\rm:} A_n)  =
\sum_{i=1}^{n} H(A_1 \cdots A_{i-1} A_{i+1} \cdots A_n)
} \hspace{3cm}\nonumber \\
&-&(n-1)\, H(A_1 \cdots A_n)
\label{mon1}
\end{eqnarray}
and
\begin{eqnarray}
\lefteqn{S_n(A_1 {\rm:} \cdots {\rm:} A_n)
= I(A_1 {\rm:} A_2 A_3 \cdots A_n)} \hspace{1cm}\nonumber \\
& & + {}  \sum_{i=2}^{n-1} I(A_i{\rm:} A_{i+1}
\cdots A_n | A_{1} \cdots A_{i-1}) \,\, , \label{rec3}
\end{eqnarray}
where $I(A{\rm:}B|C) = H(AC) + H(BC) - H(C) - H(ABC)$ is the
conditional mutual information between $A$ and $B$ given $C$.
The proof of these different equivalent expressions follows
from the following recurence relation for $S_n$:
\begin{eqnarray}
\lefteqn{S_n(A_1{\rm:} \cdots {\rm:} A_n)=
S_{n-1}(A_1{\rm:} \cdots {\rm:} A_{n-1} A_n)
} \hspace{3cm}\nonumber \\
&+& I(A_{n-1}{\rm:}A_n |A_1 \cdots A_{n-2}) \,\, .
\label{rec}
\end{eqnarray}

These expressions allow us to derive
the following simple  properties of $S_n$:
\begin{enumerate}
\item $S_n$ is symmetric under the interchange of any two parties
  $A_i$ and $A_j$. This follows from Eq. (\ref{SSS1}).
\item $S_n$ is semi-positive. This follows from
  Eq. (\ref{rec3}) and from the positivity of the
  conditional mutual entropy, $I(A{\rm:}B|C)\ge 0$,
 which itself follows from the strong
  subbaditivity of Shannon entropies.
\item $S_n$ is additive.
\item $S_n$ vanishes on product
probability distribution $P=P_{A_1}P_{A_2}\ldots P_{A_n}$.
\item For two parties $S_2$ is the mutual information
\begin{eqnarray}
S_2 (A{\rm:}B) & = & H(A) + H(B) - H(AB) \nonumber \\
& = & I (A{\rm:}B)\,\, .
\end{eqnarray}
\end{enumerate}

\subsection{Local increase in entropy to erase all correlations:
  $T_n$}

The second secrecy monotone is defined as
\begin{eqnarray}
T_n (A_1 {\rm:} \cdots {\rm:} A_n) & = &
\sum_{i=1}^{n} H(A_i) - H(A_1 \cdots A_n) \,\, . \nonumber \\
\label{mon2}
\end{eqnarray}
In order to interpret this quantity we note that it is equal to the
minimum relative entropy between the probability distribution
$P_{A_1\ldots A_n}$ and any product probability distribution
$Q_{A_1}Q_{A_2}\ldots Q_{A_n}$ (with the minimum being attained when the
$Q_{A_i}$ are equal to the local distributions $Q_{A_i}$):
\begin{eqnarray}
&T_n (A_1 {\rm:} \cdots {\rm:} A_n) \nonumber\\
& = D(P_{A_1\ldots A_n}||P_{A_1}P_{A_2}\ldots P_{A_n})&  \nonumber \\
& =  \min_{Q_{A_1}Q_{A_2}\ldots Q_{A_n}}
D(P_{A_1\ldots A_n}||Q_{A_1}Q_{A_2}\ldots Q_{A_n})
 \,\, .&
\label{t2}
\end{eqnarray}
where $D(P_A||Q_A)= \sum_a P(a)\ln {P(a)\over Q(a)}$ is the relative
entropy between the distributions $P$ and $Q$.
In order to give an interpretation to $T_n$, we turn to recent work of
Vedral\cite{Vedral} (see also the review \cite{Vedral2})
who gave an interpretation of a related quantity, the relative
entropy of entanglement, as the minimum increase of entropy of
{\em classically correlated} environments needed to erase all
correlations between the parties sharing an entangled
states. (The relative entropy of entanglement is the minimum relative
entropy between the entangled state and any separable state).
Vedral's argument can easily be extended to the present situation
whereupon one finds that $T_n$ is the minimum increase of entropy of
local {\em uncorrelated} environments if the parties erase all
correlations between them by interacting
locally with their  environment.

To proceed, we note that $T_n$ obeys the recurrence relation
\begin{eqnarray}
\lefteqn{T_n(A_1{\rm:} \cdots {\rm:} A_n)=
T_{n-1}(A_1{\rm:} \cdots {\rm:} A_{n-1})
}\hspace{3cm}\nonumber\\
&+&  I(A_n{\rm:} A_1\cdots A_{n-1})
\label{rec2}
\end{eqnarray}
which allows us to derive the following expression:
\begin{eqnarray}
\lefteqn{T_n (A_1 {\rm:} \cdots {\rm:} A_n )
= I(A_1 {\rm:} A_2)} \hspace{1cm}\nonumber\\
& & + \sum_{i=2}^{n-1} I(A_1\cdots A_i{\rm:}A_{i+1}) \,\, . \label{rec4}
\end{eqnarray}

These expressions allow us to derive
the following simple  properties of $T_n$:
\begin{enumerate}
\item $T_n$ is symmetric under the interchange of any two parties
  $A_i$ and $A_j$. This follows from Eq. (\ref{mon2}).
\item $T_n$ is semi-positive. This follows from Eq. (\ref{rec4}).
\item $T_n$ is additive.
\item $T_n$ vanishes on product
probability distribution $P=P_{A_1}P_{A_2}\ldots P_{A_n}$.
\item For two parties $T_2$ is the mutual information
\begin{eqnarray}
T_2 (A{\rm:}B) & = & H(A) + H(B) - H(AB) \nonumber \\
& = & I (A{\rm:}B)\,\, .
\end{eqnarray}
\end{enumerate}

\subsection{Relation between $S_n$ and $T_n$}

For two parties, $S_n$ and $T_n$ coincide and are equal to the mutual
entropy between the parties.
Thus, $S_n$ and $T_n$
can be viewed as two (generally distinct) multipartite extensions
of the mutual information of a bipartite system.
That these two generalizations are generally distinct follows from the
following relation between the two monotones:
\begin{eqnarray}
\lefteqn{S_n(A_1{\rm:} \cdots {\rm:} A_n)
+ T_n (A_1{\rm:} \cdots {\rm:} A_n)} \nonumber \\
 & = &\sum_{i=1}^n I(A_i {\rm:} A_1 \cdots A_{i-1} A_{i+1} \cdots A_n ) \,\, .
\label{SnTn}
\end{eqnarray}
This expression will prove important in the interpretation of the
monotones in the quantum case.

Let us note that linear combinations of $T_n$ and $S_n$ of the form
\begin{equation}
M_n = \lambda S_n + (1- \lambda) T_n  \,\, ,
\end{equation}
with $0\leq \lambda \leq 1$
are monotones as well. For the case of three parties, we will prove below that
{\em only} for this range of $\lambda$ is $M_n$ a monotone.

\subsection{Monotonicity of $S_n$ and $T_n$ under local operations}

We now prove that $S_n$ is a monotone, i.e., it can
can only decrease under LOCC.
Local operations by party $j$ correspond to  carrying out a local
transformation which modifies $A_j$ to $\bar{A}_j$ according to the
conditional probability distribution $P_{\bar{A}_j | A_j}$.
For example, let us choose $A_n$ to undergo such a transformation.
We want to prove first that
\begin{equation}
S_n(A_1 {\rm:} \cdots {\rm:} A_n) \ge S_n(A_1 {\rm:} \cdots {\rm:}
\bar{A}_n) \,\, ,
\end{equation}
Using Eqs. (\ref{rec2}) and (\ref{SnTn}), we find
\begin{eqnarray}
\lefteqn{S_n(A_1{\rm:} \cdots {\rm:} A_n)
= - T_{n-1} (A_1{\rm:} \cdots {\rm:} A_{n-1})} \nonumber \\
 & + &\sum_{i=1}^{n-1} I(A_i {\rm:} A_1 \cdots A_{i-1} A_{i+1} \cdots
A_{n-1}A_n ) \,\, .
\label{eins}
\end{eqnarray}
Clearly, only the second term on the right hand side is affected by the local
operation on $A_n$. As a consequence of the
data processing inequality (see e.g. \cite{it}), one can show that each term
of the summation can only decrease under
the transformation $A_n \to \bar{A}_n$,
\begin{eqnarray}
\lefteqn{ I(A_i {\rm:} A_1 \cdots A_{i-1} A_{i+1} \cdots A_{n-1}A_n ) }
\nonumber\\
&\ge& I(A_i {\rm:} A_1 \cdots A_{i-1} A_{i+1} \cdots A_{n-1}\bar{A}_n
) \ .
\end{eqnarray}
For example, consider the term $i=1$, and
write the mutual information $I(A_1{\rm:}A_2 \cdots A_n \bar{A}_n)$
in two equivalent ways:
\begin{eqnarray}
\lefteqn{ I(A_1{\rm:}A_2 \cdots A_{n-1}\bar{A}_n)+I(A_1{\rm:}A_n
  |A_2\cdots A_{n-1}\bar{A}_n)
} \hspace{0.5cm}\nonumber\\
& & = I(A_1{\rm:}A_2\cdots A_n)+I(A_1{\rm:}\bar{A}_n|A_2\cdots A_n)
\end{eqnarray}
We have $I(A_1{\rm:}\bar{A}_n|A_2 \cdots A_n)=0$ since $A_1$ and
$\bar{A}_n$ are conditionally independent given $A_n$.
Using strong subadditivity $I(A_1{\rm:}A_n|A_2\cdots A_{n-1}\bar{A}_n)\ge 0$,
we conclude that $I(A_1{\rm:}A_2 \cdots A_n)
\ge I(A_1{\rm:}A_2 \cdots A_{n-1} \bar{A}_n)$.
Finally, as $S_n$ is symmetric in all $A_j$, this proof is
actually valid for local operation performed by all parties.

In order to prove the monotonicity of $T_n$ under local operations,
we assume, as above, that $A_n$ undergoes a local transformation to
$\bar{A}_n$, and prove that
\begin{equation}
T_n(A_1 {\rm:} \cdots {\rm:} A_n) \ge T_n(A_1 {\rm:} \cdots {\rm:}
\bar{A}_n) \,\, .
\label{TnLO}
\end{equation}
Using Eq. (\ref{rec2}), we have
\begin{eqnarray}
\lefteqn{T_n(A_1 {\rm:} \cdots {\rm:} A_{n-1} {\rm:} \bar{A}_n)} \nonumber \\
& = & T_{n-1}(A_1 {\rm:} \cdots {\rm:} A_{n-1}) +
I(\bar{A}_n {\rm:} A_1 \cdots A_{n-1})   \,\, .
\end{eqnarray}
Again, due to the data processing inequality, the second term
on the right hand side cannot increase as a result of
the local transformation on $A_n$, while the first term remains
unchanged. This proves Eq.~(\ref{TnLO}).
Consequently, as $T_n$ is symmetric in all $A_j$'s, it can
only decrease under local operations of any party.

\subsection{Monotonicity of $S_n$ and $T_n$ under public classical communication}

Now, let us consider the monotonicity of $S_n$ and $T_n$
under classical communications.
Here, classical communication means that one party makes its
probability distribution (partly or completely)
known to all the other parties.
Say, we choose the party $A_1$ to make $\bar{A}_1$ known to the
public, where $\bar{A}_1$ is drawn from the conditional probability
distribution $P_{\bar{A}_1|A_1}$.
We want to prove that $S_n$ is a monotone, that is,
\begin{equation}
S_n(A_1 {\rm:} \cdots {\rm:} A_n)
\ge  S_n(A_1 {\rm:} \cdots {\rm:} A_n | \bar{A}_1) \,\, ,
\end{equation}
with the right hand side term being the monotone $S_n$
calculated from the probability distribution
$P_{A_1\cdots A_n|\bar{A}_1=a}$, averaged over all values $a$
of $\bar{A}_1$, or
\begin{eqnarray}
\lefteqn{S_n(A_1 {\rm:} \cdots {\rm:} A_n | \bar{A}_1)=
\sum_{i=1}^{n} H(A_1 \cdots A_{i-1} A_{i+1} \cdots A_n| \bar{A}_1)
} \hspace{4cm} \nonumber\\
&-& (n-1) \, H(A_1 \cdots A_n| \bar{A}_1) \nonumber\\
\end{eqnarray}
Using Equation~(\ref{rec3}), we have
\begin{eqnarray}
\lefteqn{S_n(A_1 {\rm:} \cdots {\rm:} A_n | \bar{A}_1)
= I(A_1 {\rm:} A_2 \cdots A_n | \bar{A}_1)} \nonumber \\
&  & + {}  \sum_{i=2}^{n-1} I(A_i{\rm:} A_{i+1}
\cdots A_n | A_{1} \cdots A_{i-1} \bar{A}_1) \,\, ,
\end{eqnarray}
The knowledge of $\bar{A}_1$ clearly only changes a conditional mutual
information if $A_1$ is {\em not} given. This is only the case in the first
term on the right hand side of the above equation.
Finally, we can prove that this term only decreases under classical
communication by writing the mutual information
$I(A_1\bar{A}_1{\rm:}A_2\cdots A_n)$ in two equivalent ways
\begin{eqnarray}
\lefteqn{
I(A_1{\rm:}A_2\cdots A_n)+I(\bar{A}_1{\rm:}A_2\cdots A_n | A_1)
}\nonumber\\
&=& I(\bar{A}_1{\rm:}A_2\cdots A_n)+I(A_1{\rm:}A_2\cdots A_n | \bar{A}_1)
\end{eqnarray}
We have $I(\bar{A}_1{\rm:}A_2\cdots A_n | A_1)=0$ since
$\bar{A}_1$ is independent of $A_2\cdots A_n$ conditionally on $A_1$.
Then, using
$I(\bar{A}_1{\rm:}A_2\cdots A_n)\ge 0$, we find that
\begin{eqnarray}
I(A_1 {\rm:} A_2 \cdots A_n) \geq I(A_1 {\rm:} A_2 \cdots A_n | \bar{A}_1)
\end{eqnarray}
which proves that $S_n$ is a monotone when party $A_1$ makes
$\bar{A}_1$ public.
Since $S_n$ is symmetric in all parties, we have also proven that it
decreases on average under classical communication between all parties.

Let us finally prove the monotonicity of $T_n$ under classical communications.
If one of the parties, say $A_1$, makes $\bar{A}_1$ public, then
$T_n$ changes according to
\begin{equation}
T_n(A_1 {\rm:} \cdots {\rm:} A_n)
\ge T_n (A_1 {\rm:} \cdots {\rm:} A_n | \bar{A}_1 )
 \,\, .
\end{equation}
with the right hand side term being the monotone $T_n$ for the
probability distribution $P_{A_1\cdots A_n|\bar{A}_1=a}$,
averaged over all values $a$ of $\bar{A}_1$, or
\begin{equation}
T_n (A_1 {\rm:} \cdots {\rm:} A_n | \bar{A}_1)
= \sum_{i=1}^n H(A_i|\bar{A}_1)
- H(A_1 \cdots A_n | \bar{A}_1)
\,\, .
\label{Tn}
\end{equation}
Using Equation~(\ref{rec4}), we have
\begin{eqnarray}
T_n (A_1 {\rm:} \cdots {\rm:} A_n | \bar{A}_1)
=\sum_{i=1}^{n-1} I(A_1\cdots A_i{\rm:}A_{i+1}|\bar{A}_1)
\end{eqnarray}
As proven above, we have for all the terms on the right hand side
\begin{equation}
I(A_1\cdots A_i{\rm:}A_{i+1}) \geq
I(A_1\cdots A_i{\rm:}A_{i+1}|\bar{A}_1)
\end{equation}
which proves that $T_n$ can only decrease under public
communication of one party. This is true for all parties
since $T_n$ is a symmetric quantity.

\section{Tripartite classical secrecy monotones}\label{tri}

\subsection{Five independent tripartite secrecy monotones}

For three parties $A$, $B$, and $C$, we have a closer look at the above
secrecy monotones for classical probability distributions.
We start by writing the
monotones explicitly in terms of entropies or mutual informations:
\begin{eqnarray}
\lefteqn{S_3 (A{\rm:}B{\rm:}C)} \nonumber \\
 & = & H(AB) + H(BC) + H(AC) - 2 H(ABC) \nonumber \\
&= &  I(A{\rm:}BC) + I(B{\rm:}C|A) \,\, ,
\\
\lefteqn{T_3 (A{\rm:}B{\rm:}C)}\nonumber \\
& = &  H(A) + H(B) + H(C) - H(ABC) \nonumber \\
& = &  I(A{\rm:}B) + I(AB{\rm:}C)  \,\, .
\end{eqnarray}

In addition to these two tripartite monotones, we
also have three other monotones $S_2(A{\rm:}BC)= I(A{\rm:}BC)$,
$S_2(B{\rm:}AC)=I(B{\rm:}AC)$ and $S_2(C{\rm:}AB)=I(C{\rm:}AB)$ 
which consist of evaluating the bipartite monotone $S_2$
on the probability distribution obtained by grouping two of the three
parties together. Thus, there is a total of 5 tri-partite secrecy
montones. These monotones are not all linearly independent as Eq. (\ref{SnTn})
shows. However, none of these monotones can be written as a linear
combination of the
other monotones with only positive coefficients. For this reason these 5
monotones give independent constraints on the transformations that are
possible under LOCC.

\subsection{Five particular probability distributions}

We begin by using these five tripartite monotones to investigate
in detail five particular tripartite probability distributions. These
five probability distributions play a particular
role since they are, in a sense made precise below, the extreme
points in a convex set.  These five distributions
consist of three bipartite distributions
\begin{eqnarray}
P^2_{AB} (0,0) & = & P^2_{AB} (1,1) = 1/2\ , \label{P21}\\
P^2_{AC} (0,0) & = & P^2_{AC} (1,1) = 1/2 \ ,\label{P22}\\
P^2_{BC} (0,0) & = & P^2_{BC} (1,1) = 1/2 \ ,\label{P23}
\end{eqnarray}
and two tripartite distributions
\begin{eqnarray}
P^3_{ABC} (0,0,0) & = & P^3_{ABC} (1,1,1) = 1/2\ ,
\label{P3}
\end{eqnarray}
and
\begin{eqnarray}
\lefteqn{
P^x_{ABC} (0,0,0) = P^x_{ABC} (1,1,0) = P^x_{ABC}(1,0,1)  }\hspace{3cm}
\nonumber \\
&& = P^x_{ABC} (0,1,1) =  1/4  \label{PXOR} \,\, .
\end{eqnarray}
The first three probability distributions, Eq.~(\ref{P21} - \ref{P23}),
are perfectly correlated shared
random bits between two of the three parties, the fourth probability
distribution, Eq.~(\ref{P3}), is one shared random bit between the
three parties, and the last
probability distribution, Eq.~(\ref{PXOR}), corresponds to the case where
two parties share an
uncorrelated probability distribution while the third party has the
exclusive-{\sc or} ({\sc xor}) of the bits of these two parties.

We can now make a table which lists for each of these probability
distributions the values of the 5 tri-partite monotones.

\begin{center}
\renewcommand{\arraystretch}{1.5}
\begin{tabular}{|c||c|c|c|c|c|}
\hline
 & $S_2(A\!\!:\!\!BC)$ & $S_2(B\!\!:\!\!AC)$ & $S_2(C\!\!:\!\!AB)$ &
$S_3(ABC)$ & $T_3(ABC)$ \\
\hline \hline
$P^2_{AB}$ & 1 & 1 & 0 & 1 & 1 \\
\hline
$P^2_{AC}$ & 1 & 0 & 1 & 1 & 1 \\
\hline
$P^2_{BC}$ & 0 & 1 & 1 & 1 & 1 \\
\hline
$P^3_{ABC}$ & 1 & 1 & 1 & 1 & 2 \\
\hline
$P^x_{ABC}$ & 1 & 1 & 1 & 2 & 1 \\
\hline
\end{tabular}
\end{center}

\subsection{Converting a probability distribution into another}

We can use this table to study which probability distributions can be
converted into which others, and with what yield.
The first thing we note from the table is that it forbids the conversion of
a probability distribution $P^x_{ABC}$ into a probability distribution
$P^3_{ABC}$ and vice-versa, as $S_3(P^x_{ABC}) > S_3 (P^3_{ABC})$ and
$T_3(P^3_{ABC}) > T_3( P^x_{ABC})$. This can be understood in the
following way. The number of shared random bits underlying the
distribution $P^x_{ABC}$ is 2 (two parties must have uncorrelated
random bits) while it is only 1 for the distribution
$P^3_{ABC}$ (where the three parties share one common bit). Since
the number of shared bits $S_3$
is a monotone, one cannot go from $P^3_{ABC}$ to $P^x_{ABC}$.
On the other hand, the number of bits that must be forgotten
in order to get three independent bits is equal to 2 for the
distribution $P^3_{ABC}$ (two parties, say  $B$ and $C$,
must randomize their bits), while it is only 1 for the distribution
$P^x_{ABC}$ (where it is enough that party C forgets its bit
in order to get independent bits). Since the number of bits
that must be forgotten to get independent distributions $T_3$ is a monotone,
one cannot go from $P^x_{ABC}$ to $P^3_{ABC}$.

The above table also suggests that distillation
procedures of the form ${P^x_{ABC}}\rightarrow P^2_{AB}$ or
${P^{3}_{ABC}}\rightarrow P^2_{AB}$ are possible. This is indeed the case:
starting from $P^x_{ABC}$, the party $C$ simply has to make its bit
public in order to get $P^2_{AB}$, thereby reducing by one
the number of shared bits $S_3$.
If we start with $P^3_{ABC}$ instead,
the party $C$ has to forget its bit, i.e., send it
through a channel which completely randomizes it. Thus, one bit
must be forgotten, reducing by one the monotone $T_3$.

The transformations ${P^3_{ABC}}^{\otimes 2} \rightarrow P^x_{ABC}$
and ${P^x_{ABC}}^{\otimes 2} \rightarrow P^3_{ABC}$
are also allowed by the above table of monotones, and we can check that
they can actually be achieved.
If the probability distribution is ${P^3_{ABC}}^{\otimes 2}$
and the parties want to have instead $P^x_{ABC}$, then $A$ has to forget
the first of the two bits it has, $B$ has to forget the second, and $C$ just
takes the sum of the two bits it has, forgetting the individual values.
Thus, three bits must be forgotten, reducing the value of $T_3$
from 4 to 1.
To get from ${P^x_{ABC}}^{\otimes 2}$ to $P^3_{ABC}$ is a little bit more
complicated. We start with $A$ having the bits $x$ and $x^\prime$, $B$ having
the bits $y$ and $y^\prime$ and $C$ having the bits $x+y$ and $x^\prime +
y^\prime$. Now $A$ makes $x$ public and $B$ makes $y^\prime$ public. From this
$C$ can calculate $y$ as well as $x^\prime$. Then, C makes
$y+x^\prime$ public, which allows
$A$ (who still has $x^\prime$) to calculate $y$. Thus, every party knows
the secret bit $y$, so we have got $P^3_{ABC}$. Here, 3 bits must have
been made public, reducing the value of $S_3$ from 4 to 1.

The above table leaves open the question whether the conversion
$$P^x_{ABC} \otimes  P^3_{ABC} \rightleftharpoons P^2_{AB}
\otimes P^2_{BC}\otimes  P^2_{AC}$$
is possible. We have not been able
to devise a protocol that carries out this transformation. 
Ruling out this possibility would
probably require an additional independent monotone, and the five
monotones listed above are the only ones we know at present.

Let us note that in order to carry out the above conversions, we
sometimes had to suppose that one of the parties forgets some of his
information. In practice, this is obviously a stupid thing to do. Why
to forget something you know? However, there may be an accident,
say an irrecoverable hard disk crash, such that one of the parties has
lost part or all of his data. In this case, the monotone $T_n$ constrains how
much secrecy is left among the parties.
It would be interesting and important to study the restricted class of
transformations in which the parties never forget their data (they would
only be allowed to communicate classically). This would impose another
constraint on the transformations that are possible.

\subsection{Extremality of the five tri-partite probability distributions}

The above discussion
raises the general question of the {\em reversible} conversion of
one probability distribution into another. By this we mean that, in
the limit of a large number of draws, it is possible 
to go from one probability distribution $P_1$ to
another $P_2$ and back with negligible losses. In particular, in the
tripartite case, one can inquire whether their are yields
$y_1,\ldots,y_5$ such that the reversible conversion
\begin{equation}
P_{ABC}\rightleftharpoons P^{2\otimes y_1}_{AB}
\otimes P^{2\otimes y_2}_{BC}\otimes  P^{2\otimes y_3}_{AC}
\otimes P^{x\otimes y_4}_{ABC} \otimes  P^{3\otimes y_5}_{ABC}
\label{39}
\end{equation}
is possible? Let us show that the five
secrecy monotones introduced above leave open the possibility of
the reversible distillation of Eq. (\ref{39}). Whether this
is possible in practice is an open question.

\begin{figure}[h]
\centering
\resizebox{7cm}{!}{\includegraphics{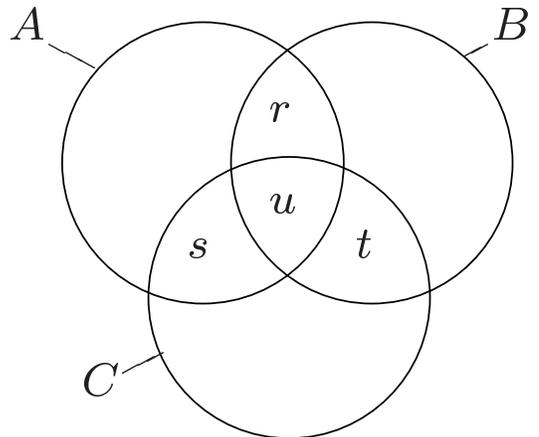}}
\caption{Venn diagram for a tripartite probability distribution.}
\label{diagram}
\end{figure}

To prove this, let us introduce the following notation:
\begin{eqnarray}
r&=& I(A{\rm:}B|C)\ ,\nonumber\\
s&=& I(B{\rm:}C|A)\ ,\nonumber\\
t&=& I(C{\rm:}A|B)\ ,\nonumber\\
u&=& I(A{\rm:}B) - I(A{\rm:}B|C)\ .
\end{eqnarray}
Let us note that $u$ is symmetric between the three parties and can
also be written as
$u =I(B{\rm:}C) - I(B{\rm:}C|A)=I(C{\rm:}A) - I(C{\rm:}A|B)$. 
These different quantities can
be represented graphically as in Fig.~\ref{diagram}.

Given these quantities we can express $S_3$ and $T_3$ as
\begin{eqnarray}
S_3  & = & r + s + t + u
\\
T_3  & = & r + s + t + 2 u  \,\, .
\end{eqnarray}
We note that $u$ is not positive definite, but we have the
positivity conditions
\begin{eqnarray}
r&\geq& 0\ ,\nonumber\\
s&\geq& 0\ ,\nonumber\\
t&\geq& 0\ ,\nonumber\\
r+u&\geq& 0\ ,\nonumber\\
s+u&\geq& 0\ ,\nonumber\\
t+u&\geq& 0\ .
\end{eqnarray}
Using these conditions, one can show that if
$u=0$, then the reversible conversion
$$P_{ABC}\rightleftharpoons P^{2\otimes y_1}_{AB}
\otimes P^{2\otimes y_2}_{BC}\otimes  P^{2\otimes y_3}_{AC}
$$
is allowed by our tripartite monotones. If $u>0$, then the
reversible conversion
$$P_{ABC}\rightleftharpoons P^{2\otimes y_1}_{AB}
\otimes P^{2\otimes y_2}_{BC}\otimes  P^{2\otimes y_3}_{AC}
\otimes  P^{3\otimes y_5}_{ABC}
$$
is allowed by our tripartite monotones. If $u<0$, then the
reversible conversion
$$P_{ABC}\rightleftharpoons P^{2\otimes y_1}_{AB}
\otimes P^{2\otimes y_2}_{BC}\otimes  P^{2\otimes y_3}_{AC}
\otimes P^{x\otimes y_4}_{ABC}
$$
is allowed by our tripartite monotones.

Thus our monotones in principle allow the reversible
conversion between any
tripartite probability distribution and the distributions
$P^{2}_{AB}$, $P^{2}_{BC}$,  $P^{2}_{AC}$,
$P^{x}_{ABC}$, and $P^{3}_{ABC}$. Whether or not such a reversible
transformation is possible or not is an open question. To rule this out
will probably require discovering additional secrecy monotones.

\subsection{Extremality of the monotones $S_3$ and $T_3$}

As a final comment about the secrecy monotones in the tripartite case,
we note that using the distillation procedures for
${P^x_{ABC}} \rightarrow P^2_{AB}$ and ${P^{3}_{ABC}} \rightarrow P^2_{AB}$,
we can now also prove that $0 \leq \lambda\leq 1$ is the only
range for which the linear combination of $S_3$ and $T_3$ is a monotone.
This can be seen by calculating $M_3 = \lambda S_3 + (1 - \lambda) T_3$
for both distillations. In the first case,
we get that $M_3(P^x_{ABC})= \lambda + 1$ should be greater
or equal to $M_3 (P^2_{AB}) = 1$, so that $\lambda \geq 0$.
In the second case, we find that
$M_3 (P^3_{ABC}) = 2 -  \lambda \geq M_3 (P^2_{AB}) = 1$,
so that $\lambda \leq 1$.
This suggests that if there are other monotones than the
$M_n$'s, they will probably not be composed out of entropies.

\section{Quantum multipartite secrecy monotones}\label{quantum}

\subsection{Definition of Quantum Secrecy Monotones}

The definition of classical secrecy monotone of Section \ref{deff} can
be immediately extended to the quantum case. The monotone will now be a
function defined on multi-partite density matrices $\rho_{A_1\ldots A_n}$
which must be:
\begin{itemize}
\item positive,
\item vanishing on product density matrices,
\item monotonous under local operations (local CP maps),
\item monotonous under classical communication,
\item additive,
\item continous.
\end{itemize}

One can also extend the quantum definition of the secrecy monotone to the case
where there is an eavesdropper. In that case, it is defined on a
multipartite density matrix $\rho_{A_1\ldots  A_n E}$. 
The monotonicity properties are then modified to require
that the secrecy montone is monotonically decreasing under local
operations and public communication by the parties $A_1\ldots A_n$ and
monotonically increasing under local operations and public communication by Eve.

In what follows, we shall for simplicity not include Eve in the
discussion. That is, we shall suppose that initially Eve has no
information about the density matrix, but she listens to all
public communications and thereby tries to thwart the parties
$A_1\ldots A_n$.

\subsection{Quantum version of the secrecy monotones $S_n$ and $T_n$}

The definitions of the  monotones
$S_n$ and $T_n$, Eqs.~(\ref{mon1}) and (\ref{mon2}),
have straightforward generalizations to the quantum case:
\begin{eqnarray}
\lefteqn{S_n (\hat{\rho}_{A_1 \cdots A_n}) }\nonumber \\
& \equiv & S(A_1 {\rm:} \cdots {\rm:} A_n)\nonumber\\
& = & \sum_{i=1}^n S(\hat{\rho}_{A_1 \cdots A_{i-1} A_{i+1} \cdots A_n})
- (n-1) S(\hat{\rho}_{A_1 \cdots A_n})
\label{qmon1}
\end{eqnarray}
and
\begin{eqnarray}
\lefteqn{T_n (\hat{\rho}_{A_1 \cdots A_n}) }\nonumber\\
& \equiv & T(A_1 {\rm:} \cdots {\rm:} A_n) \nonumber\\
& = &  \sum_{i=1}^n S(\hat{\rho}_{A_i} ) 
- S(\hat{\rho}_{A_1 \cdots A_n}) \,\,
,
\label{qmon2}
\end{eqnarray}
where now $S(\hat{\rho})$ denotes the von Neumann entropy of a density matrix
which is given by $S(\hat{\rho}) = - {\mathrm{Tr}}(\hat{\rho} \log
\hat{\rho})$ and partial traces are written in the form $\hat{\rho}_{A_1
\cdots A_{i-1} A_{i+1} \cdots A_n} = {\mathrm{Tr}}_{A_i} (\hat{\rho}_{A_1
\cdots A_n})$.

The different rewritings of $S_n$ [Eqs. (\ref{SSS1}), (\ref{rec3}), and
(\ref{rec})] and $T_n$ [Eqs. (\ref{rec2}) and(\ref{rec4})]
that where obtained in the classical case carry through to
the quantum case, in analogy to the what was shown for bipartite 
systems in \cite{ca2}.  This means that the simple properties
that followed from these rewritings in the classical case also hold
in the quantum case. In particular, the positivity of the $S_n$ and
$T_n$ follows from the positivity of the conditional mutual entropy,
which holds in both the classical and quantum case (see \cite{Wehrl}
for a review). The proofs of monotonicity change in the quantum case,
and we give them below.

Let us note that, for pure states, $S_n$ and $T_n$ coincide and are
equal to the sum of the local entropies:
\begin{eqnarray}
S_n (|\psi_{A_1 \cdots A_n}\rangle)
&=& T_n (|\psi_{A_1 \cdots A_n}\rangle)\nonumber \\
& = &  \sum_{i=1}^n S(\hat{\rho}_{A_i}) \ .
\label{pure}
\end{eqnarray}
Thus for instance on a singlet state, $S_2$ and $T_2$ are equal to 2,
and on a GHZ state, $S_3$ and $T_3$ are equal to 3.

We do not at present have a clear interpretation of $S_n$ in the
quantum case. On the other hand, the interpretation of $T_n$ in the
quantum case is the same as in the classical case.
Indeed, it can be written as the minimum relative entropy between
$\hat\rho_{A_1\ldots A_n}$ and a product density matrix
$\hat\eta_{A_1}\otimes\ldots \otimes\hat\eta_{A_n}$
(the minimum being attained when $\hat\eta_{A_i}=\hat\rho_{A_i}$).
Therefore, $T_n$ can be interpreted as the minimum increase of entropy of
local (uncorrelated) environnements if the parties erase all
correlations between them by letting their quantum systems interact
with a local environment.

\subsection{Monotonicity of $S_n$ and $T_n$}

We now give the proofs of monotonicity of $S_n$ and $T_n$ under local
operations and classical communication in the quantum case.

Local operations of one party are described mathematically as completely
positive (CP) local maps $M_{A_i}$, which only act on the subspace of the
$i$th party. We can assume that such a map is implemented as follows
\cite{CPmap,Kraus}:
$A_i$ adds to its Hilbert space an auxiliary variable in a pure
state $\Pi_{\mathrm{aux}} = |0\rangle_{\mathrm{aux}}\langle 0 |$. It then
carries out a unitary transformation $\hat{U}_{i \mathrm{aux}}$ on its
original system and the auxiliary variable. Finally, it traces over a part
{\it aux}$^\prime$ of her Hilbert space. Note that {\it aux}$^\prime$
does not have to coincide with {\it aux}. Hence, we can represent a local
CP map as
\begin{eqnarray}
\tilde{\rho}_{A_1 \cdots A_n} & = & M_{A_i} \otimes
\openone_{A_1 \cdots A_{i-1} A_{i+1} \cdots A_n} (\hat{\rho}_{A_1 \cdots A_i})
\nonumber \\
& = & {\mathrm{Tr}}_{\mathrm{aux}^\prime} \left[
(\hat{U}_{A_i {\mathrm{aux}}} \otimes
\openone_{A_1 \cdots A_{i-1} A_{i+1} \cdots A_n})\right. \nonumber \\
&  & \left. (\hat{\rho}_{A_1 \cdots A_n}
\otimes \Pi_{\mathrm{aux}} ) \right. \nonumber \\
&  & \left.( \hat{U}^\dagger_{A_i \mathrm{aux}} \otimes
\openone_{A_1 \cdots A_{i-1} A_{i+1} \cdots A_n}) \right] \,\, .
\label{CPmap}
\end{eqnarray}

We start with $S_n$ and write it in the following form
\begin{eqnarray}
\lefteqn{S_n(
%\hat{\rho}_{A_1 \cdots  A_n}; 
A_1 {\rm:} \cdots {\rm:} A_n)   } \nonumber \\
& = &  \sum_{i=1}^{n-1} S_2(
%\hat{\rho}_{A_1 \cdots A_n}; 
A_i {\rm:} A_1 \cdots A_{i-1} A_{i+1} \cdots A_{n-1} A_n )\nonumber \\
&  &  - {}  \Delta_n \,\, ,
\label{S2}
\end{eqnarray}
with
\begin{equation}
\Delta_n = \sum_{i=1}^{n-1} S(\hat{\rho}_{A_i} ) - S(\hat{\rho}_{A_1 \cdots
A_{n-1}}) \,\,
.
\end{equation}
Now we assume that the system $A_n$ undergoes a local CP map $M_n$,
Eq.~(\ref{CPmap}). As $\Delta_n$ does not depend on $A_n$ it remains
unchanged, thus we only have to check $S_2$ for monotonicity. For this we
rewrite Eq.~(\ref{CPmap}) for two systems $A$ and $B$
\begin{eqnarray}
\tilde{\rho}_{AB} &=& M_A  \otimes \openone_B (\rho_{AB} )\nonumber\\
&=&
{\mathrm{Tr}}_{a'}( U_{Aa} \otimes \openone_B )
\rho_{AB} \otimes \Pi_a ( U_{Aa}^\dagger \otimes \openone_B )
\end{eqnarray}
and note that neither adding a local auxiliary nor performing a unitary
transformation changes $S_2$. Tracing over a local subsystem, however,
decreases $S_2$ since
\begin{eqnarray}
\lefteqn{S_2(
%\hat{\rho}_{A^\prime {\mathrm{aux}}^\prime  B }; 
A^\prime {\mathrm{aux}}^\prime {\rm:} B) - S_2 (
%\hat{\rho}_{A^\prime  B};
A^\prime {\rm:} B )} \nonumber \\
&  = & S (
%\hat{\rho}_{A^\prime  BA }; 
{\mathrm{aux}}^\prime {\rm:} B | A^\prime) \,\,
,  \hspace*{3cm}
\end{eqnarray}
which is just the conditional mutual quantum entropy and which, due to
strong subadditivity \cite{Wehrl} is semipositive, thus implying that
\begin{equation}
S_2(
%\hat{\rho}_{A^\prime {\mathrm{aux}}^\prime  B }; 
A^\prime {\mathrm{aux}}^\prime {\rm:} B ) \geq  S_2 (
%\hat{\rho}_{A^\prime  B}; 
A^\prime {\rm:} B )\,\, .
\end{equation}
Due to symmetry, $S_n$ given by Eq.~(\ref{S2}) is then monotone under local CP
maps of any party.

For monotonicity under local measurements and public communication of their
outcome, we assume that a positive operator valued measurement (POVM)
\cite{Kraus} is performed on system $A_1$. This is realized by adding as above
an ancilla $\Pi_{\mathrm{aux}}$ to $A_1$ and then carrying out a von Neumann
measurement that transforms $\hat{\rho}_{A_1 \cdots A_n}\otimes
\Pi_{\mathrm{aux}}$ to
\begin{eqnarray}
\tilde{\rho}_{{\mathrm{aux}} A_1 \cdots A_n}  =  \sum_k  \hat{\rho}^k_{
{\mathrm{aux}} A_1 \cdots A_n}
 =  \sum_k p_k \tilde{\rho}^k_{{\mathrm{aux}} A_1 \cdots A_n}\,\, ,
\end{eqnarray}
with $ \hat{\rho}^k_{{\mathrm{aux}} A_1 \cdots A_n} = (\hat{P}^k_{
{\mathrm{aux}} A_1} \otimes \openone_{A_2 \cdots A_n})
 ( \hat{\rho}_{A_1 \cdots
A_n} \otimes \Pi_{\mathrm{aux}} )(\hat{P}^k_{{\mathrm{aux}} A_1} \otimes
\openone_{A_2 \cdots A_n} )$ and $\hat{P}^k_{ {\mathrm{aux}} A_1}$ a complete
set of orthogonal projectors acting on the extended space
${\mathrm{aux}} A_1$, $\tilde{\rho}^k_{{\mathrm{aux}}A_1 \cdots A_n}$ being
the joint state after outcome $k$ has been measured and $p_k = {\mathrm{Tr}}
(\hat{\rho}^k_{{\mathrm{aux}}A_1 \cdots A_n})$. We now go back to
Eq.~(\ref{qmon1}). 
The orthogonality of the projectors $\hat{P}^k_{{\mathrm{aux}} A_1 }$
implies that the $\tilde{\rho}_{A_1 \cdots A_{i-1} A_{i+1} \cdots A_n}$ are
block diagonal for $i \neq 1$, so that their entropies can be expressed as
\begin{eqnarray}
S(\tilde{\rho}_{A_1 \cdots A_{i-1} A_{i+1} \cdots A_n}) &  =  & H[ p_k]
\nonumber \\
&  & + {} \sum_k
p_k S (\tilde{\rho}^k_{A_1 \cdots A_{i-1} A_{i+1} \cdots A_n}) \,\, ,
\nonumber \\
\label{cond1}
\end{eqnarray}
and
\begin{eqnarray}
S(\tilde{\rho}_{A_1 \cdots A_n}) & = & H[p_k] + \sum_k p_k
S(\tilde{\rho}^k_{A_1 \cdots A_n}) \,\, ,
\label{cond2}
\end{eqnarray}
with $H[p_k]$ denoting the classical Shannon entropy of the probability
distribution $p_k$.
For $i=1$, we find the following inequality for the first term, which makes use
of the concavity of entropy
\begin{eqnarray}
S( \tilde{\rho}_{A_2 \cdots A_n}) \geq \sum_k p_k S(\tilde{\rho}^k_{A_2 \cdots
A_n}) \,\, .
\label{conc}
\end{eqnarray}
Replacing all these expressions in Eq.~(\ref{mon1}), we finally find that
\begin{eqnarray}
S_n \left(\sum_k p_k \tilde\rho_{A_1\cdots A_n}^k\right) \geq
\sum_k p_k  S_n(\tilde\rho_{A_1\cdots A_n}^k) \,\, .
\label{convexSn}
\end{eqnarray}
This shows that the monotone $S_n$ can only decrease on average if $A_1$
performs a POVM measurement and the outcome is made known to the other parties. 
By symmetry, this property holds for all $A_i, i= \{1, \ldots, n\}$.

To prove the monotonicity of $T_n$ we proceed as follows. Suppose that
$A_1$ carries out a local CP map. As before adding a local ancilla
 and carrying out a local unitary transformation
do not change $T_n$. Tracing over part of $A_1$'s Hilbert space
decreases $T_n$. Indeed,
$T_n(
%\hat{\rho}_{A_1  a_1\cdots A_n}; 
A_1 a_1 {\rm:} A_2 {\rm:} \cdots {\rm:} A_n ) -
T_n (
%\hat{\rho}_{A_1\cdots A_n}; 
A_1 {\rm:} \cdots {\rm:} A_n) =
S(a_1 {\rm:} A_2\ldots A_n | A_1) \geq 0$.
Suppose now that $A_1$ carries out a measurement (with outcomes $k$)
and publicly reveals
the result. In Eq.~(\ref{qmon2}),  the terms $S(\rho_{A_i})$ with $i\neq 1$
decrease because of concavity of entropy [see Eq.~(\ref{conc})] and 
because the term  $S(\rho_{A_1}) -
S(\rho_{A_1\ldots A_n})$ stays constant [where we used Eqs.~(\ref{cond1}) and
(\ref{cond2})].  Hence,
\begin{eqnarray}
\label{convexityTn}
\lefteqn{T_n\left(\sum_k p_k \tilde{\rho}_{A_1\cdots A_n}^k; A_1 : \cdots : A_n
\right)} \nonumber \\ &  \geq &
\sum_k p_k  T_n(\tilde{\rho}_{A_1\cdots A_n}^k; A_1 : \cdots : A_n)
\end{eqnarray}
where we have used the same notation as in Eq.~(\ref{convexSn}).

\subsection{Applications of quantum secrecy monotones}

The two quantum monotones described above can be used to provide
bounds on the rate of conversion of one multipartite density matrix
into another using local operations and classical communication.
As an example, we study in this section and the next one how many
realizations of a correlated tripartite probability distributions 
can be obtained from  a GHZ state.

Let us recall that the GHZ state, in the $z$ basis, is
$$|GHZ\rangle = (|000\rangle +
|111\rangle )/\sqrt{2}\ .$$
If the state is measured in the $z$ basis one obtains the probability
distribution $P^3$. In contrast, if the state is measured in the $x$ basis 
one obtains the probability distribution $P^x$.

We have shown above that $P^3$ and $P^x$ cannot be
reversibly converted one into the other. This therefore suggests that
when  using a GHZ state
to do multipartite quantum cryptography, there is an irreversible choice
that must be made. However, the above discussion leaves open the
possibility that the three parties could
use a more sophisticated strategy than the ones just described
and thereby
obtain more than one of these probability distributions from a single GHZ
state.

To address this question, let us compute the monotones $S_3$ and $T_3$
on the initial state and on the final probability distributions. We
find
\begin{eqnarray}
S_3(|GHZ\rangle)=3\quad &,&\quad T_3(|GHZ\rangle)=3\ , \nonumber\\
S_3(P^3)=1\quad &,&\quad T_3(P^3)=2\ , \nonumber\\
S_3(P^x)=2\quad &,&\quad T_3(P^x)=1\ .
\label{STSTST}
\end{eqnarray}
Thus the monotones leave open the possibility of a higher yield than
one $P^3$ or one $P^x$ per GHZ state.

Let us note however an interesting feature of eq. (\ref{STSTST}),
namely that the sum of the final values of $S_3$ and $T_3$ is equal to
half the sum of the initial values:
\begin{eqnarray}
S_3(P^3) + T_3(P^3)&=&S_3(P^x)+ T_3(P^x)
\nonumber\\
&=&{
S_3(|GHZ\rangle)+T_3(|GHZ\rangle) \over 2} \ .
\label{Sum}
\end{eqnarray}
We shall now show that this is no accident but is necessarily the case
when one passes from a multipartite pure state to a multipartite
probability distribution. Thus it is indeed impossible to obtain
more than one $P^3$ or one $P^x$ probability distribution from a
single GHZ state, and the simple measurement strategies described
above are therefore optimal.

\subsection{Decrease of $S_n+T_n$ when passing from a multipartite
pure state to a multipartite probability distribution}

Let us suppose that initially the parties share a multipartite {\em pure}
state
$|\Psi_{A_1\ldots A_n}\rangle$. Initially
\begin{eqnarray}
S_n(\|Psi_{A_1\ldots A_n}\rangle)=T_n(|\Psi_{A_1\ldots A_n}\rangle)=
\sum_i S(\rho_{A_i}) \ .
\label{In}
\end{eqnarray}
Suppose that the aim of the parties is to obtain, by carrying out
local measurements and classical communication, a multipartite
probability distribution $P_{A_1\ldots A_n}$. In doing so, the
monotones $S_n$ and $T_n$ will decrease. More precisely,
the amount by which they decrease 
is such that their sum is decreased by at least a factor two:
\begin{eqnarray}
S_n(P_{A_1\ldots A_n})+T_n(P_{A_1\ldots A_n})\leq
\sum_i S(\rho_{A_i}) \ .
\label{Out}
\end{eqnarray}

To prove this, let us first consider the bipartite case. Thus
initially the parties share a pure state $|\Psi_{AB}\rangle$ and they
carry out measurements so as to obtain a probability distribution
$P_{AB}$. Let us first suppose that no communication takes place
between the parties. Then, it follows from Holevo's
bound\cite{Holevo} that the mutual information between Alice and
Bob after the measurement is necessarily less than the local
entropies of the original state:
\begin{eqnarray}
S(\rho_A)=S(\rho_B) \geq I_{P_{AB}}(A{\rm:}B)\ .
\label{SSI}
\end{eqnarray}
Equality is attained in Eq. (\ref{SSI}) only if they measure in
the Schmidt basis.

Let us now show that Eq. (\ref{SSI}) also holds if the parties
communicate classically.  We will suppose that the communication
takes place in a series of rounds. During each round, one of the
parties carries out a partial measurement on the state and
communicates information to the other party. After all the
communication has taken place the parties measure the state they
are left with. Such a general protocol is difficult to analyze, but we can
transform it into a simpler protocol. In the simpler protocol,
during each round the party transmits all the information obtained
by the partial measurement to the other party. This should be
contrasted with the most general protocol in which only part of
the information obtained by the measurement is transmitted. The
simplification follows from the fact that we can divide the
measurement into a first partial measurement in which the
information transmitted to the other party is obtained, and a
second partial measurement in which the information that is kept
is obtained. But the second partial measurement could then as well
be carried out during the next round. Repeating this reasoning
round after round, we can construct a simpler protocol in which
the information that is not communicated to the other party is
acquired during the last round only.

In the case of the simplified protocol, one can easily show that
Eq. (\ref{SSI}) holds. Consider the first round. Suppose that
Alice carries out a partial measurement. The measurement has
outcomes $k$, with probabilities $p(k)$. The state if the outcome
is $k$ is $\Psi_{AB}^k$. Because of monotonicity of the quantum
mutual information, we have
\begin{equation}
S(\rho_A) \geq \sum_k p(k) S(\rho_A^k)\ .
\end{equation}
The local entropies decrease (on average) due to the
communication. The same will hold for all the subsequent rounds.
Hence, Eq. (\ref{SSI}) holds also if the parties carry out public
communication. In fact that above reasoning shows that the optimal
strategy is for the parties not to communicate, but simply to
measure the state in the Schmidt basis.

Finally let us consider the multipartite case. The result for 
two parties Eq.(\ref{SSI}) implies
that for any partition of the parties into one party, say
$i$, and $n-1$ parties, the mutual information between $i$ and the $n-1$
other parties after the measurements is bounded by
\begin{eqnarray}
I(A_i{\rm:}A_1\ldots A_{i-1}A_{i+1}\ldots A_n)\leq S(\rho(A_i)) \  .
\label{InS}
\end{eqnarray}
Summing over $i$ and using Eq. (\ref{SnTn}), we find that
\begin{eqnarray}
S_n(P)+T_n(P)\leq \sum_i S(\rho(A_i))
\label{STS}
\end{eqnarray}
which is what we wanted to prove.

\section{Conclusion}

In this article, we have introduced the concept of secrecy monotones
which are powerful tools to obtain bounds on the distillation rate 
in Maurer's classical cryptographic scheme 
as well as bounds on the distillation rate in quantum cryptography.

We introduced two independent multipartite secrecy monotones 
based on (Shannon or von Neuman) entropies, $S_n$ and $T_n$,
which allowed us to investigate the distillation rates for
multipartite cryptographic schemes. In the classical case, we studied
in detail the tripartite case and showed that their are several
inequivalent tripartite probability distributions in the sense that
they cannot be converted reversibly one into the other.
We also studied the particular case of tripartite quantum
cryptography based on the GHZ state. We showed that
the parties must choose a priori which probability distribution they want
to generate.  

The important feature that emerges from our study is thus that in
multipartite classical or quantum cryptography, the parties must make
an irreversible choice on what final probability distribution they
want to obtain. Making the wrong choice entails an irreversible loss. 
We note that this feature is not unique to
cryptography; indeed, a similar situation arises in multipartite
entanglement distillation since there are entangled pure states that
cannot be reversibly converted one into the other\cite{BPRST,LPSW}.

{\bf Note:} After this paper was completed, we learned of the work
\cite{MMM} in which monotones (under certain classes of
operations) which are positive both on quantum states and on
probability distributions are considered in the bipartite case.

{\bf Acknowledgements:} We would like to thank Nicolas Gisin and
Daniel Collins for helpful conversations. We acknowledge funding 
by the European  Union
under the project EQUIP (IST-FET programme). S.M. is a research associate
of the Belgian National Fund for Scientific Research.

\end{document}